\input psfig.sty
\documentclass[referee]{aa}
\thesaurus{07.16.1}
\title{Numerical tests of dynamical friction in gravitational inhomogeneous systems}
\author{A. Del Popolo\inst{1,2,3} 
}
\offprints{A. Del Popolo - {\it E-mail adelpop@unibg.it}}
\institute{$^1$ Dipartimento di Matematica, Universit\`{a} Statale
di Bergamo,
via dei Caniana, 2 - I 24129 Bergamo, ITALY \\
$^2$ Feza G\"ursey Institute, P.O. Box 6 \c Cengelk\"oy, Istanbul, Turkey \\
$^3$ Bo$\breve{g}azi$\c{c}i University, Physics Department,
80815 Bebek, Istanbul, Turkey\\
}
\titlerunning{Numerical tests of dynamical friction}
\authorrunning{A. Del Popolo}
\date{Received 28 January 2003, Accepted 10 April 2003}

\begin{document}
\maketitle

\label{firstpage}

\begin{abstract}

In this paper, I test by numerical simulations the results of Del Popolo \& Gambera (1998),
dealing with the extension of Chandrasekhar and von Neumann's analysis of the
statistics of the gravitational field to systems in which particles
({\it e.g.} stars, galaxies) are inhomogeneously distributed.
The paper is an extension of that of Ahmad \& Cohen (1974), in which the authors 
tested some results of the stochastic theory of dynamical friction developed by 
Chandrasekhar \& von Neumann (1943) in the case of homogeneous gravitational systems.
It is also a continuation of the work developed in 
Del Popolo (1996a,b), which extended the results of Ahmad \& Cohen (1973), 
(dealing with the study of the probability distribution of the stochastic force in homogeneous gravitational
systems) to inhomogeneous gravitational systems. 
%
%%In Del Popolo \& Gambera (1998), we derived a
%%distribution function $ W({\bf F},d{\bf F}/dt)$ giving the joint
%%probability that a test
%%particle is subject to a force {\bf F} and an associated rate of change of
%%{\bf F} given by d{\bf F}/dt. We calculated the first moment of d{\bf F}/dt
%%to study the effects of inhomogenity on dynamical friction.
%
Similarly to what was done by Ahmad \& Cohen (1974) in the case of homogeneous systems, 
I test, by means of the evolution of an inhomogeneous 
system of particles,
% (Plummer's sphere), 
%the stochastic force in inhomogeneous gravitational 
%systems
%using N-body realizations of Plummer's spherically symmetric 
%models, 
that the theoretical rate of force fluctuation d{\bf F}/dt describes correctly
the experimental one, 
I find that the stochastic force distribution obtained for 
the evolved system is in good agreement with the Del Popolo \& Gambera (1998) theory.
Moreover, in an inhomogeneous
background the friction force is actually enhanced relative to the
homogeneous case.

\end{abstract}

%
%%\keywords{stars: statistics - galaxies: star clusters-cluster of-
%%interactions - cosmology: large scale structure of Universe}
%

\section{Introduction}

The study of the statistics of the fluctuating gravitational force in infinite 
homogeneous systems was pioneered by Chandrasekhar \& Von Neumann in two classical papers 
(Chandrasekhar \& Von Neumann 1942, 1943 hereafter CN43) and in several other papers by Chandrasekhar 
(1941, 1943a, 1943b, 1943c, 1943d, 1943e, 1944a and 1944b). The analysis of
the fluctuating gravitational field, developed by the authors, was
formulated by means of a statistical treatment. In their papers 
Chandrasekhar \& Von Neumann considered a system in which the
stars are distributed according to a uniform probability density, 
no correlation among the positions of the stars is present and 
where the number of stars constituting the system tends to infinity while
keeping the density constant.
%homogeneous system a system made of stars but their results are more 
%general and they can be extend to systems of galaxies or of clusters 
%of galaxies.
Two distributions are fundamental for the description of the fluctuating 
gravitational field: \\
1) $ W({\bf F})$ which gives the probability that a test star
is subject to a force ${\bf F}$ in the range ${\bf F}$, ${\bf F}$+
d${\bf F}$;\\
2) $ W({\bf F},{\bf f})$ which gives the
joint probability that the star experiences a force {\bf F} and a rate 
of change {\bf f}, where $ {\bf f} = d{\bf F}/dt $.\\
The first distribution, known as Holtsmark's law (Holtsmark 1919), in the 
case of a homogeneous distribution of the stars, gives information only 
on the number of stars experiencing a given force but it does not 
describe some fundamental features of the fluctuations in 
the gravitational field such as the {\it speed of the fluctuations} and 
the dynamical friction. These features can be 
described using the second distribution $ W({\bf F},{\bf f})$.
Hence, for the definition of the speed of fluctuations and of the 
dynamical friction one must determine the distribution $ W({\bf F},{\bf f})$. 
Information on dynamical friction can be obtained from the moments of this last distribution.  
As stressed by Chandrasekhar \& von Neumann (1943), for a test star moving with velocity {\bf v} in
a sea of field stars characterized by a random probability distribution 
of the velocities, $ \Phi({\bf u})$, we may write:
\begin{equation}
\langle {\bf V} \rangle = \langle {\bf u} \rangle - {\bf v} = - {\bf v}
\label{eq:tr}
\end{equation}
where ${\bf V}$ represents the velocity of a typical field star relative
to the one under consideration, ${\bf u}$ denotes the velocity of a
field star. 
This asymmetry of the distribution of the relative velocities produces, as
shown by CN43, a
deceleration of the test star in the direction of motion.
This effect is known, from the Chandrasekhar papers, as  ``{\it dynamical friction}".
As quoted above, some information on dynamical friction can be obtained by means of the 
first moment of {\bf f}. As shown by CN43:
\begin{equation}
\langle \frac{d{\bf F}}{dt} \rangle_{{\bf F},{\bf v}} = \frac{-2 \pi}{3} G m n  B(\beta) \left[ {\bf v} - \frac{3 {\bf F} \cdot {\bf v}}{ | {\bf F} |^2} \cdot {\bf F} \right]
\label{eq:qu}
\end{equation}
\begin{equation}
\langle \frac{d | {\bf F} |}{dt} \rangle_{{\bf F},{\bf v}} = \frac{ 4 \pi}{3} G m n  B(\beta) \frac{{\bf F} \cdot {\bf v}}{ | {\bf F} |}
\label{eq:cin}
\end{equation}
where $ m$ is the mass of a field star, $ n_{l}$ is the local density, and 
$ B(\beta)$ is definited 
as:
\begin{equation}
B\left( \beta \right) =3\frac{\int_{0}^{\beta }H(\beta )d\beta }{\beta H\left( \beta \right) }-1
\end{equation}
where
\begin{equation}
H\left( \beta \right) =\frac{2}{\pi \beta }\int_{0}^{\infty }e^{-\left( \frac{x}{\beta }\right) ^{\frac{3}{2}}}x\sin (x)dx
\end{equation}
and $ \beta = | {\bf F}| / Q_{H}$ 
%= | {\bf F}| / 2.603 G M n^{2/3}$.
where $Q_{H}=\left(\frac{4}{15}\right)^{2/3} 2 \pi G M n^{2/3}$.
%in CN43 Eq. (98).
These equations show that the amount of acceleration in the direction 
of $- {\bf v}$ when $ {\bf v} \cdot {\bf F} \le 0$ is greater than that 
in the direction $+ {\bf v}$, when $ {\bf v} \cdot {\bf F} \ge 0$: then  
the star suffers a deceleration, the a priori probability that 
$ {\bf v} \cdot {\bf F} \ge 0$ being equal to the probability that 
$ {\bf v} \cdot {\bf F} \le 0$. \\
%Being equal the a priori probability to have  $ {\bf v} \cdot {\bf F} \ge 0$ 
%or $ {\bf v} \cdot {\bf F} \le 0$, the star suffers a deceleration.\\
%%%%%%
%Within the context of the Chandrasekhar-von Newmann stochastic formalism
%the most serious unanswered question concerns the possibility of a
%rigorous derivation of dynamical friction.
%It has not yet proven possible to pass from the rate of change of force
%per unit mass (Eq. (\ref{eq:qu})), to an estimate of the force per
%unit mass, $ {\bf F}$, acting upon a moving star. 
%An alternative way to proceede, suggested by Chandrasekhar (1943),
%is the calculation of the autocorrelation
%function $ W({\bf F}_0,{\bf F}_t)$, giving the probability that at same 
%time  $ t = 0$, a test star experiences a force $ {\bf F}$ and at $ t$ the 
%force $ {\bf F}$. It is then possible to obtain the first momentum  
%$ \langle {\bf F}_t ({\bf F}_0, {\bf v}) \rangle $ and finally the integral  
%$ \int_0^{\infty} \langle {\bf F}_t ({\bf F}_0, {\bf v}) \rangle dt $ 
%should give dynamical friction starting from the statistical  theory.
Several authors have stressed
the importance of stochastic forces and in particular dynamical friction
in determining the observed
properties of clusters of galaxies (White 1976; Kashlinsky 1986,1987)
while others studied the role of dynamical friction in the orbital decay
of a satellite moving around a galaxy or in the merging
scenario (Bontekoe \& van Albada 1987; Seguin \& Dupraz 1996;
Dominguez-Tenreiro \& Gomez-Flechoso 1998). 
%
%%which is not only
%%the framework for galaxy formation picture in hierarchical cosmological models, but also
%%important for the study of particular aspects of the evolution of a number of astronomical
%%systems, such as galactic nuclei, cD galaxies in rich galaxy clusters.
%
Chandrasekhar's theory (and in particular his
classical formula (see Chandrasekhar 1943b))
is widely employed to quantify dynamical friction
in a variety of situations, even if
%in most astronomical problems
%the background is neither infinite nor homogeneous. It is evident
the theory developed is based on the
hypothesis that the stars are distributed uniformly and 
%%and the velocity
%%are characterized by a spherical distribution.
it is well known that in stellar systems, the stars are not 
uniformly distributed,
(Elson et al. 1987; Wybo \& Dejonghe 1996; Zwart et al. 1997) 
in galactic systems as well, the galaxies are not uniformly distributed 
(Peebles 1980; Bahcall \& Soneira 1983; Sarazin 1988; Liddle, \& Lyth 1993; 
White et al. 1993; Strauss \& Willick 1995).
%Several authors stressed
%the importance of stochastic forces and in particular dynamical friction
%in determining the observed
%properties of clusters of galaxies (White 1976; Kashlinsky 1986,1987)
%while others studied the role of dynamical friction in the orbit decaying
%of a satellite moving around a galaxie or in the merging
%scenario (Bontekoe \& van Albada 1987; Seguin \& Dupraz 1996;
%Dominguez-Tenreiro \& Gomez-Flechoso 1998) which is not only
%the framework for galaxy
%formation picture in hierarchical cosmological models, but also for more
%particular aspects of the evolution of a number of astronomical
%systems, such as galactic nuclei, cD galaxies in rich galaxy clusters.
%Chandrasekhar's theory is widely employed to quantify dynamical friction
%in a variety of situations, even if in most astronomical problems
%the background is neither infinite nor homogeneous.
It is evident
that an analysis of dynamical friction taking account of the
inhomogeneity of astronomical systems can provide a more realistic
representation of the evolution of these systems.
%%Moreover it
%%is known that Chandrasekhar's formula suffers from several drawbacks
%%that arise from the physical assumptions made in its derivation and
%%furthermore it cannot describe some situations, as for example the drag
%%experienced by a satellite placed outside the edge of a finite gravitating
%%system.
Moreover from a pure theoretical standpoint we expect that inhomogeneity affects all the aspects of the fluctuating 
gravitational field (Antonuccio \& Colafrancesco 1994;  
Del Popolo \& Gambera 1996, 1997; Del Popolo et al. 1996; Gambera 1997). 
Firstly, the Holtsmark distribution is no longer correct 
for inhomogeneous systems. For these systems, as shown by Kandrup 
(1980a, 1980b, 1983), the Holtsmark distribution must be substituted 
with a generalized form of the Holtsmark distribution
%When inhomogeneity increases the distribution of stochastic force
characterized by a shift of $W({\bf F})$
towards larger forces when inhomogeneity increases. This result 
was already suggested by the numerical simulations of Ahmad \& Cohen 
(1973, 1974). 
%and the maximum decreases in height.
Hence when the inhomogeneity increases
the probability that a test particle experiences a
large force increases, secondly, 
$ W({\bf F},{\bf f}) $ is changed by inhomogeneity. 
Consequently, the values of the mean life of a state, 
the first moment of $ {\bf f}$ and the dynamical friction force
%As a consequence dynamical friction should
%increase with increasing inhomogeneity.\\ 
are changed by inhomogeneity with respect to those of homogeneous systems.\\
This paper must be intended as the continuation of Del Popolo \& Gambera (1998) paper, in 
which 
%first part of a work
%pointed to:  \\
%a) 
%was performed 
the 
study of the effects of inhomogeneity on the distribution functions
of the stochastic forces and on dynamical friction was performed. \\
As anticipated in Del Popolo \& Gambera (1998), the next task to perform 
was to test the result of the Del Popolo \& Gambera (1998) paper 
against N-body simulations, which is the object of the present paper. 
The third step (to be developed in a future paper) should be that of 
finding a formula that describes dynamical friction in homogeneous and 
inhomogeneous systems only on the basis of
statistical theory.\\
Before continuing we want
to stress that when we speak of inhomogeneity we refer to inhomogeneity
in position distribution and not to that of velocity distribution. Our
work  follows the spirit of Kandrup's (1980) in the sense that
we are interested in
the effect of a non-uniform distribution in the position of stars on the
distributions of the stochastic force.
%We expect that inhomogeneity affects all the aspects of the fluctuating 
%gravitational field (Antonuccio \& Colafrancesco 1994; Del Popolo 1994; 
%Del Popolo \& Gambera 1996, 1997; Del Popolo et al. 1996; Gambera 1997). 
%Firstly the Holtsmark distribution is no more correct 
%for inhomogeneous systems. For these systems, as shown by Kandrup 
%(1980a, 1980b, 1983), the Holtsmark distribution must be substituted 
%with a generalized form of the Holtsmark distribution
%When inhomogeneity increases the distribution of stochastic force
%characterized by a shift of $W({\bf F})$
%towards larger forces when inhomogeneity increases. This result 
%was already suggested by the numerical simulations of Ahmad \& Cohen 
%(1973, 1974). 
%and the maximum decreases in height.
%Hence when the inhomogeneity increases
%the probability that a test particle experiences a
%large force increases, secondly 
%$ W({\bf F},{\bf f}) $ is changed by inhomogeneity. 
%Consequently, the values of the mean life of a state, 
%the first moment of $ {\bf f}$ and the dynamical friction force
%As a consequence dynamical friction should
%increase with increasing inhomogeneity.\\ 
%are changed by inhomogeneity with respect to these of homogeneous systems.\\

The plan of the paper is the following: in Sec. ~2 we review the
calculations and formulas needed to obtain $\langle \frac{d{\bf F}}{dt} \rangle_{{\bf F},{\bf v}}$
%distribution function $ W({\bf F},{\bf f}) $ after having released 
in the case of inhomogeneous systems.
In Sec. ~3, I show how numerical experiments are performed and 
they are compared with the theoretical results of Del Popolo \& Gambera (1998).
%we calculate the first
%moment of {\bf f} and in Sec. ~ 4 we 
%show how dynamical friction is influenced by inhomogeneity. 
Finally, in Sec. 4, I draw my conclusions.

%\section{Test of $\langle \frac{d{\bf F}}{dt} \rangle_{{\bf F},{\bf v}}$ in inhomogeneous systems}

%\subsection{$\langle \frac{d{\bf F}}{dt} \rangle_{{\bf F},{\bf v}}$ and dynamical friction}
%\section{$\langle \frac{d{\bf F}}{dt} \rangle_{{\bf F},{\bf v}}$ and dynamical friction}
\section{Force derivative and dynamical friction}

The introduction of the notion of dynamical friction is due to CN43. In the 
stochastic formalism developed by CN43 the dynamical friction is discussed 
in terms of $ {\bf f}$:
\begin{equation}
\overline {\bf f} =  \frac{-2 \pi}{3} G m n  B(\beta) \left[ {\bf v} - \frac{3 {\bf F} \cdot {\bf v}}{ | {\bf F} |^2} \cdot {\bf F} \right]
\label{eq:quarot}
\end{equation}
where
\begin{equation}
B( \beta) =  \frac{3 \cdot \int_{0}^{\beta} W(\beta) d \beta }{ \beta \cdot W(\beta)} \; - \; 1
\label{eq:quarno}
\end{equation}

%In
%the large force limit we have that
%\begin{equation}
%B( \beta) = \frac{8}{5} \sqrt{\frac{\pi}{2}} \beta^{1.5}
%\label{eq:cinq}
%\end{equation}
%the strong force limit can be obtained in a simple two body calculation:
%\begin{equation}
%\frac{d F}{dt} \cong F \cdot \frac{v}{R} \cong G M n \beta^{1.5} v
%\label{eq:cinquno}
%\end{equation}
%where the distance between the two body is $ R = (G M / F)^{0.5}$. The 
%calculation of $ {\bf F}$ from $ d{\bf F}/dt$ has not yet proven possible.\\
%We may only say that if 
%\begin{equation}
%T = \frac{G M}{\langle v^2 \rangle^{1.5}}
%\label{eq:cinqdue}
%\end{equation}
%is the time-scale for a close encounter then:
%\begin{equation}
%F \cong \frac{d F}{dt} \cdot T
%\label{eq:cinqtre}
%\end{equation}
As shown by CN43, the origin of dynamical friction is due to 
the asymmetry in the distribution of relative velocities. 
%As previously told, 
If a test star moves with velocity $ {\bf v}$ in a spherical 
distribution of field stars, namely $ \phi( \bf u)$, then we have that:
\begin{equation}
\overline{\bf V} = \overline{{\bf u} - {\bf v}} = - {\bf v}
\label{eq:cinqqua}
\end{equation}
The asymmetry in the distribution of relative velocities is conserved 
in the final Eq. (\ref{eq:quarot}). In fact from Eq. (\ref{eq:quarot}) 
we have:
\begin{equation}
\langle \frac{d |{\bf F}|}{dt} \rangle= \frac{4 \pi}{3} G M n B(\beta) \cdot \frac{{\bf v F}}{{\bf F}} \label{eq:cinqcin}
\end{equation}
(CN43). This means that when $ {\bf v} \cdot {\bf F} \ge 0$ then 
$ \frac{d |{\bf F}|}{dt} \ge 0$, 
while when $ {\bf v} \cdot {\bf F} \le 0$ then $ \frac{d |{\bf F}|}{dt} \le 0$.
%As a conseguence the acceleration of a star when
%$ {\bf v} \cdot {\bf F} \ge 0$ is directed as $ - {\bf v}$ while when 
%$ {\bf v} \cdot {\bf F} \le 0$ is directed as $ + {\bf v}$.\\
As a consequence, when ${\bf F}$ has a positive component in the direction of
${\bf v}$,  ${|\bf F|}$ increases on average; while if  ${\bf F}$ has
a negative component in the direction of  ${\bf v}$,  ${|\bf F|}$
decreases on average. 
%The modulus of acceleration in the first case is larger than in the second. 
%Being equal the probability to find $ {\bf v} \cdot {\bf F} \ge 0$ 
%or $ {\bf v} \cdot {\bf F} \le 0$ the test star is decelerated.
Moreover, the star suffers a greater amount of acceleration in the
direction $ - {\bf v}$ when
$ {\bf v} \cdot {\bf F} \le 0$ than in the direction $ +{\bf v}$ when 
$ {\bf v} \cdot {\bf F} \ge 0$.\\ 
In other
words the test star suffers, statistically, an equal number of 
accelerating and decelerating impulses. The modulus of 
deceleration being larger than that of acceleration the star slows down.\\

Chandrasekhar \& von Neumann's analysis was extended to inhomogeneous systems 
in Del Popolo \& Gambera (1998), where $ W({\bf F},{\bf f})$ and its first momentum were calculated. 
Supposing that the distribution function $\tau$ is given by:
\begin{equation}
\tau = \frac{a}{r^p} \psi (j^2(M) |{\bf u}|^2)
%%\frac{1}{(2 \pi \sigma^2)^\frac{3}{2}}
%%\exp{\left(-\frac{v^2}{2 \sigma^2}\right)}
\label{eq:un}
\end{equation}
where $ a$ is a constant
that can be obtained from the normalization condition for  $\tau$, $j$ a
parameter (of dimensions of velocity$^{-1}$), $\psi$ an arbitrary
function, $\bf u$ the velocity of a field star.
In other words we assumed, according to CN43 and Chandrasekhar
\& von Newmann (1942), that the distribution of velocities is spherical,
i.e. the distribution function is 
$\psi({\bf u}) \equiv  \psi (j^2(M) |{\bf u}|^2) $,
%Maxwellian (as stressed by Chandrasekhar \& von Newmann 1942, the
%results obtained remain valid for any spherical distri)
but differently from the quoted papers we supposed that the positions
are not equally likely for stars, that is, the stars are
inhomogeneously distributed in space.
A lengthy calculation led us (see Del Popolo \& Gambera 1998 for a derivation and the
meaning of symbols) to find that the first moment of  $ {\bf f}$ is given by:
\begin{equation}
\overline {\bf f} =  -\left( \frac{1}{2} \right)^{\frac{3}{3 - p}} \cdot A(p) \cdot B(p)^{\frac{p}{3 - p}} \cdot \frac {\alpha^{\frac{3}{3 - p}} G M L(\beta)}{ \pi H(\beta) \beta^{\frac{2 - p}{2}}} \cdot \left[ {\bf v} - \frac{3 {\bf F} \cdot {\bf v}}{ |  {
\bf F} |^2} \cdot {\bf F} \right]
\label{eq:quaruno}
\end{equation}
where:
\begin{equation}
A(p) =  \int_{0}^{\infty} \left[  \frac{\sin{x}}{x^{(4 - p)/2}} -  \frac{3 \sin{x}}{x^{(8 - p)/2}} + \frac{3 \cos{x}}{x^{(6 - p)/2}} \right] \cdot d x  
\label{eq:venqua}
\end{equation}

\begin{equation}
B(p) = \int_{0}^{\infty} \frac{z - \sin{z}}{z^{(7 - p)/2}} \cdot d z
\end{equation}

\begin{eqnarray}
L ( \beta) & = & 6 \int_{0}^{\infty} \left[
e^{(x/\beta)^{\frac{(3 - p)}{2}}} \right]
\left[ \frac{ \sin{x}}{x^{(2 - p)/2}} -
\frac{\cos{x}}{x^{p/2}} \right] dx \nonumber \\
           &   & - \; 2 \int_{0}^{\infty}
           \left[ e^{(x/\beta)^{\frac{(3 - p)}{2}}}
           \right] \cdot \frac{\sin{x}}{x^{(p - 2)/2}} dx
\label{eq:quardu}
\end{eqnarray}

\begin{equation}
H(\beta) = \frac{2}{\pi \beta}
\int_{0}^{\infty} e^{-\left(\frac{x}{\beta}\right)^{3/2}}
\cdot x sin(x) dx
\end{equation}

The results obtained by us for an inhomogeneous system 
are different [see Eq. (\ref{eq:quaruno})], as expected,
from those obtained by CN43 for
a homogeneous system (CN43 - Eq. 105 or Eq. (\ref{eq:quarot})). At the same time 
it is very interesting
to note that for $ p = 0$ (homogeneous system) our result coincides,
as is obvious, with the
results obtained by CN43. 
In inhomogeneous systems, Eq. (\ref{eq:cinqcin}) can be written, using
Eq. (\ref{eq:quaruno}), as:
\begin{equation}
\langle \frac{d |{\bf F}|}{dt} \rangle=  2^{\frac{p}{p-3}} \cdot A(p) \cdot B(p)^{\frac{p}{3 - p}} \cdot \frac {\alpha^{\frac{3}{3 - p}} G M L(\beta)}{ \pi H(\beta) \beta^{\frac{2 - p}{2}}} \cdot \frac{{\bf v F}}{{\bf F}}
\label{eq:q1}
\end{equation}
In order to check the validity of the quoted relation (Eq. (\ref{eq:q1})), I have performed numerical experiments.
This was done similarly to Del Popolo (1996b) by evolving (now) 100000 points (stars) acting under their mutual gravitational attraction. From the evolved positions and velocities of the stars, $\langle \frac{d {\bf F}}{dt} \rangle$ was computed as a function of velocity and force, similarly to Ahmad \& Cohen (1974), and then compared with Eq. (\ref{eq:q1}) as I shall describe in the following. 
\footnote{In a inhomogeneous system, similarly
way to what happens in a homogeneous system, $ {\bf f}$ depends
on $ {\bf v}$, $ {\bf F}$ and $ \theta$ (the angle between $ {\bf v}$ and 
$ {\bf F}$) while unlike homogeneous systems, $ {\bf f}$ 
is a function of the inhomogeneity parameter
$p$. The dependence of $ {\bf f}$ on $p$ is not only due to the
functions $A(p)$, $ B(p)$ and to the density parameter $ \alpha$ but
also to the parameter $ \beta=|\bf F|/Q_{H}$. In fact in
inhomogeneous systems the {\it normal} field $Q_{H}$ is given by
$Q_{H}=GM(\alpha B(p)/2)^{2/(3-p)}$, clearly dependent on $p$.}

%\subsection{N-body experiments}
\section{N-body experiments}

%IMPORTANTE: CAMBIARE TUTTO, IL PROFILO E' $1/r^p$

To calculate the stochastic force in an inhomogeneous system, 
I used an initial configuration in which particles were 
distributed according 
%to a Plummer model
to a truncated power-law 
density profile:  
\begin{equation}
\rho(r)=\rho_0 \left(\frac{r_0}{r}\right)^p  \hspace*{1cm}  0\leq r\leq R
\end{equation}
(see Kandrup 1980, Del Popolo 1996a). 
If the velocity distribution is everywhere isotropic then the equation relating the 
configuration space density $\rho(r)$ to the phase space density f(E) is:
\begin{equation}
\rho(r)=4\pi \int_{U(r)}^{0} \sqrt{2\left[E-U(r)\right]} f(E) dE
\label{eq:confff}
\end{equation}
where $U(r)$ is the potential (normalized to zero at infinity). Eq. (\ref{eq:confff})
may be converted into an Abel integral equation and inverted, giving
the phase space density:
%can be theoretically calculated by 
%means of:
\begin{equation}
f(E)=\frac{\sqrt{2}}{4 \pi^2} \frac{d}{dE} \int_E^0 
\frac{dU}{\sqrt{U-E}}
\frac{d \rho}{dU}
\label{eq:abell}
\end{equation}
(Eddington 1916; Binney \& Tremaine 1987).
\begin{figure}
\label{Fig. 1}
\psfig{figure=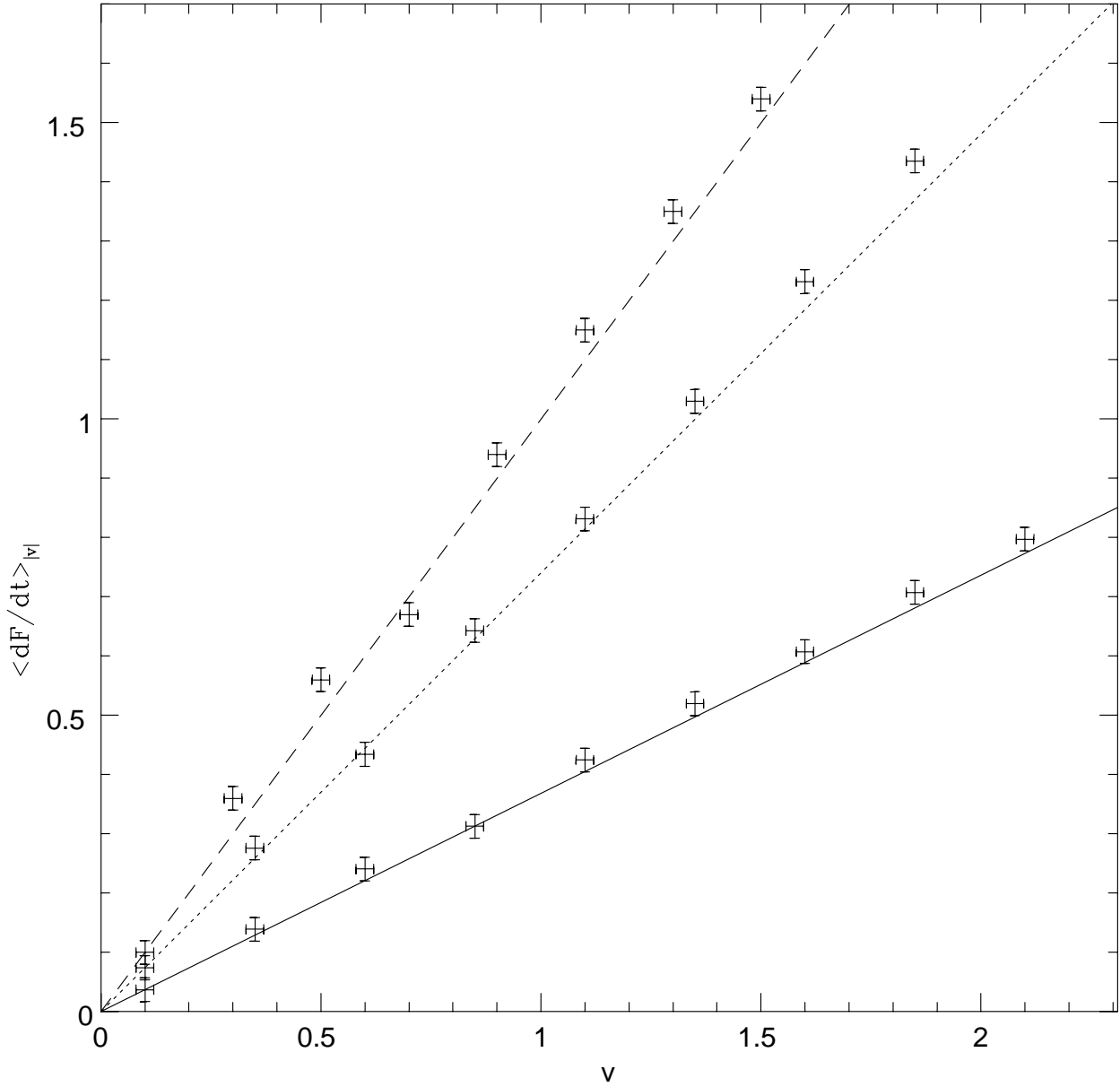,width=18cm} 
\caption[]{The average value of the time rate of change of the magnitude of the force as the function of the velocity.
The solid line refers to the homogeneous case (Chandrasekhar \& von Neumann 1943). The dotted and dashed line refers to the cases $p=2.5$ and $p=4$, respectively (see Eq. 26). Crosses represent the experimental points.}
\end{figure}
The initial conditions were generated 
from the distribution function that can be obtained from 
%given in 
Eq.~(\ref{eq:abell})
assuming a cut-off radius $ R=1$, the mass of the system $ M=1$, 
$ r_{0} =0.15$ \footnote{This is the value I used, remember however that 
the distribution is scale-free} and $ G=1$. All the particles had  
equal mass.  
To have a system whose total mass is contained in a unitary 
sphere, Eq. (\ref{eq:confff})
was renormalized and 
consequently also the potential of the system 
which is obtained from Eq. (\ref{eq:confff})
through Poisson's equation. 
The system of 100000 particles 
was evolved over 150 dynamical times using a tree N-body code 
(Hernquist 1987). 
%This N-body code was modified to calculate 
%the total force on a test point at the centre of the system. 
During the evolution of the system the essential quantities such as position, force, etc.,  
%total force 
%acting on 
for each test point of the system was sampled 
every $ \frac{1}{20}$ of a dynamical time (see Del Popolo 1996b, Ahamad \& Cohen 1973 for details). 
\begin{figure}
\label{Fig. 1}
\psfig{figure=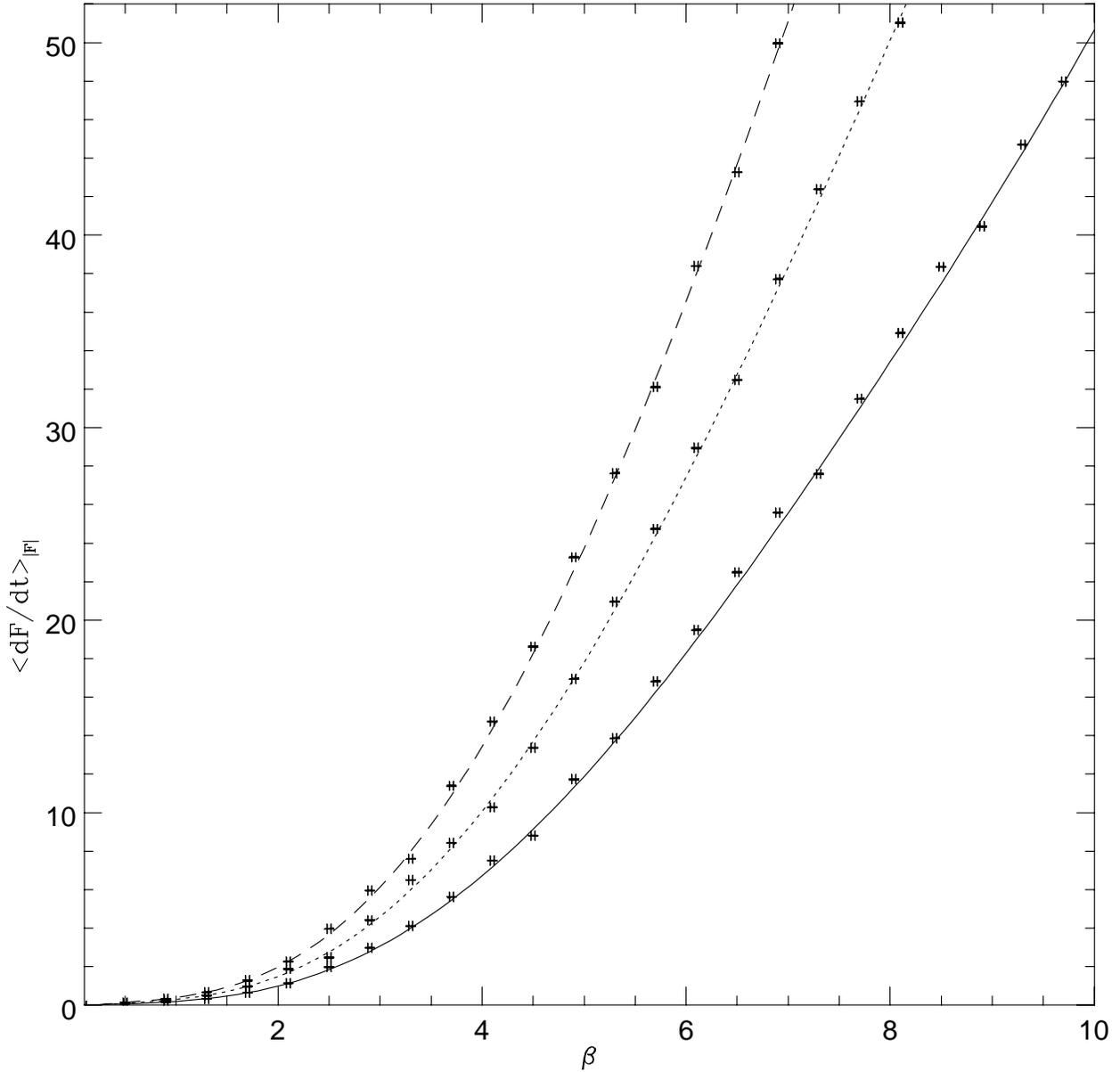,width=18cm} 
\caption[]{The average value of the time rate of change of the magnitude of the force as the function of the force.
The solid line refers to the homogeneous case (Chandrasekhar \& von Neumann 1943). The dotted and dashed line refers to the cases $p=2.5$ and $p=4$, respectively (see Eq. 19). Crosses represent the experimental points.}
\end{figure}
The average of the $ \frac{d |{\bf F}|}{dt}$ is a function of velocity, $v$, force, $F$, 
and the angle between them. The test of Eq. (\ref{eq:q1}) was performed in a similar way to that of 
Ahmad \& Cohen (1974), namely by integrating out two of the variables and examining
$ \langle \frac{d |{\bf F}|}{dt} \rangle$ against the remaining one (see Ahmad \& Cohen 1974).
As in Ahmad \& Cohen (1974), $ \langle \frac{d |{\bf F}|}{dt} \rangle_{|{\bf F}|}$ indicates $ \langle \frac{d |{\bf F}|}{dt} \rangle$ after integrating out the angle and velocity, while $ \langle \frac{d |{\bf F}|}{dt} \rangle_{|{\bf v}|}$ is 
$ \langle \frac{d |{\bf F}|}{dt} \rangle$ after integrating out the angle and force. In integrating out the angle, 
one cannot average over the entire range 0 to $\pi$, since that would give zero. 
Instead, the cosine was averaged from 0 to $\pi$/2, and in the numerical experiments only those particles having a cosine in the quoted range were used. In order not to waste the statistics for half the particles the same trick of Ahmad \& Cohen (1974) was used, namely when the cosine is in the range of $\pi/2$ to $\pi$, the sign of $ \frac{d |{\bf F}|}{dt}$ is changed and it is counted in the same statistics. This corrensponds to assume that the cosine between ${\bf v}$ and ${\bf F}$ is uniformly distributed, which is what is found in numerical experiments. The average value of the cosine between 0 and $\pi/2$ is $1/2$. 

%If one assunes 
For a general distribution $\langle |{\bf v}| \rangle$ can be calculated as usual:
\begin{equation}
\langle |{\bf v}| \rangle=\int_0^{\infty}\frac{f(E) {\bf v} d{\bf v}}{f(E)d{\bf v}}
\end{equation}
%and Eq (\ref{eq:q1}) 
and $\langle \frac{d |{\bf F}|}{dt} \rangle$
can be written, in units of $\frac{\left(\frac{1}{2}\right)^{\frac{3}{3-p}}}{\pi} \alpha^{\frac{3}{3 - p}} G M \langle |{\bf v}| \rangle$, as:
\begin{equation}
\langle \frac{d |{\bf F}|}{dt} \rangle_{|{\bf F}|}=
A(p) \cdot B(p)^{\frac{p}{3 - p}} \cdot \frac { L(\beta)}{ H(\beta) \beta^{\frac{2 - p}{2}}} \label{eq:q11}
\end{equation}
%$\frac{2^{\frac{p}{p-3}}}{\pi} \alpha^{\frac{3}{3 - p}} G M \langle |{\bf v}| \rangle$ 
In the particular case of 
a Maxwellian distribution for velocities:
\begin{equation}
\psi= \frac{j^3}{\pi^{3/2}} exp(-j^2 |{\bf v}|^2)
\end{equation}
where $j^2=\frac{3}{2 <{\bf v}^2>}$, so that:
\begin{equation}
\langle |{\bf v}| \rangle=\frac{2}{\pi^{1/2}j}
\end{equation}
we have that:
\begin{equation}
\langle \frac{d |{\bf F}|}{dt} \rangle_{|{\bf F}|}=  \frac{2^{\frac{p}{p-3}}}{\pi^{3/2}j} \cdot A(p) \cdot B(p)^{\frac{p}{3 - p}} \cdot \frac {\alpha^{\frac{3}{3 - p}} G M L(\beta)}{ H(\beta) \beta^{\frac{2 - p}{2}}} 
%\cdot \frac{{\bf v F}}{{\bf F}}
\label{eq:q2}
\end{equation}
that expressed in units of 
$\frac{2^{\frac{p}{p-3}}}{\pi^{3/2}j} \alpha^{\frac{3}{3 - p}} G M$ 
, then Eq. (\ref{eq:q2}) becomes:
\begin{equation}
\langle \frac{d |{\bf F}|}{dt} \rangle_{|{\bf F}|}=  A(p) B(p)^{\frac{p}{3 - p}} \cdot \frac { L(\beta)}{ H(\beta) \beta^{\frac{2 - p}{2}}} 
%\cdot \frac{{\bf v F}}{{\bf F}}
\label{eq:q3}
\end{equation}
 
Similarly to Ahamd \& Cohen (1974), since to integrate out the force from Eq. (\ref{eq:q1}) one has a divergent result, I consider only particles up to a certain maximum value of the force, $\beta_{\rm max}$: for example in the case $p=0$
$\beta_{\rm max}=10.7$, that involves 97\% of the particles. As observed by Ahmad \& Cohen (1974), any cutoff of the force can be used as long as it is taken into account in both the experiment and the analytic evaluation of $\langle \frac{d |{\bf F}|}{dt} \rangle_{|{\bf v}|}$. 
%So, similarly to Ahmad \& Cohen (1974) while   
Defining:
\begin{equation}
B_1(\beta)=\frac { L(\beta)}{\pi H(\beta) \beta^{\frac{2 - p}{2}}} 
\end{equation}
and 
\begin{equation}
\langle B_1(\beta) \rangle=\frac{\int_0^{\beta_{\rm max}} B_1(\beta) H(\beta) d\beta}{\int_0^{\beta_{\rm max}} H(\beta) d\beta}
\label{eq:med}
\end{equation}
we find, in units of $\left(\frac{1}{2}\right)^{\frac{3}{p-3}} \langle B1(\beta)\rangle \alpha^{\frac{3}{3 - p}} G M$,
that:
%
%%We have that:
%%\begin{equation}
%%\langle \frac{d |{\bf F}|}{dt} \rangle=2^{\frac{p}{p-3}} 
%%\cdot A(p) \cdot B(p)^{\frac{p}{3 - p}} \cdot 
%%\alpha^{\frac{3}{3 - p}} G M 
%%\langle B_1(\beta) \rangle v
%%\label{eq:q4}
%%\end{equation}
%and in units of $2^{\frac{3}{p-3}} \langle B1(\beta)\rangle \alpha^{\frac{3}{3 - p}} G M$
%and in units of $\left(\frac{1}{2}\right)^{\frac{3}{p-3}} \langle B1(\beta)\rangle \alpha^{\frac{3}{3 - p}} G M$
%we have that:
\begin{equation}
\langle \frac{d |{\bf F}|}{dt} \rangle_{|{\bf v}|}=
%2^{\frac{p}{p-3}} 
A(p) \cdot B(p)^{\frac{p}{3 - p}} \cdot 
%\alpha^{\frac{3}{3 - p}} G M 
%L(\beta)}{ \pi H(\beta) \beta^{\frac{2 - p}{2}}} 
%\langle B1(\beta) \rangle 
v
%\cdot \frac{{\bf v F}}{{\bf F}}
\label{eq:q5}
\end{equation}
%\section{Results}

The results of calculation and numerical experiments are plotted in Fig. 1 and Fig. 2.
In Fig. 1, I plot the average value of the time rate of change of the magnitude of the force as a function of the velocity. The solid line refers to the homogeneous case while the dotted and dashed lines refer to the cases $p=2.5$ and $p=4$, respectively. Crosses represent the experimental points. 
As shown, experimental points follow a linear relationship and there is a good agreement with the theoretical prediction, (Eq. \ref{eq:q11}).
In Fig. 2, I plot the average value of the time rate of change of the magnitude of the force as the function of the force.
As in the previous figure, the solid line refers to the homogeneous case while the dotted and dashed line refers to the cases $p=2.5$ and $p=4$, respectively. In this case, the dependence is no longer linear: it behaves like $B(\beta)$ in the homogeneous case.
A comparison with numerical experiments shows that there is a good agreement with the theoretical prediction, (Eq. \ref{eq:q5}).
The situation described in Ahmad \& Cohen (1974), 
%namely the fact 
that the experimental data were somewhere in between the theoretical curves for the one-particle and the infinite-particle case, is no longer present and the agreement is better, now. This is due to the larger number of particles used in the simulations. The plots show that Chandrasekhar \& von Neumann's theory of dynamical friction in gravitational systems gives a good description of experimental data (solid line and data), and so does the generalization of the quoted theory to inhomogeneous systems (dotted line, dashed line, and respective data). In inhomogeneous systems, Chandrasekhar's result 
which relates the frictional force only to the local properties of the background at the position of the object, is no
longer true, and friction depends on the global structure of the system.  
This point is in agreement with Maoz (1993), who showed that in inhomogeneous media the friction, 
unlike Chandrasekhar's formula, depends on the global structure of the entire mass density field.

\section{Conclusions}

In this paper, I tested by numerical simulations the results of the Del Popolo \& Gambera (1998) paper,
dealing with the average value of the time rate of change of the magnitude of the stochastic force in inhomogeneous gravitational systems. In agreement with Ahmad \& Cohen (1974), 
the stochastic theory of dynamical friction developed by 
Chandrasekhar \& von Neumann (1943), in the case of homogeneous gravitational systems, gives a good description of the results of numerical experiments.
The stochastic force distribution obtained for 
inhomogeneous systems, obtained by Del Popolo \& Gambera (1998), is also 
in good agreement with the results of numerical experiments.
Finally, in an inhomogeneous
background the friction force is actually enhanced relative to the
homogeneous case.

\section*{Acknowledgments}
%%%%%%Work partially supported by funds ex-60\% 98.
%%V.A.-D. would like to thank Prof. Giuseppe Moncada.
I am grateful to E. N. Ercan and G. Mamon for stimulating discussions during
the period in which this work was performed. I would
like to thank Bo$\breve{g}azi$\c{c}i University Research
Foundation for the financial support through the project code
01B304.

\end{document}